\documentclass[prd,aps,showpacs,nofootinbib]{revtex4}%
\usepackage{amssymb}
\usepackage{graphicx}
\usepackage{amsmath}
\usepackage{amsfonts}%
\newcommand{\be}{\begin{equation}}
\newcommand{\ee}{\end{equation}}
\newcommand{\bq}{\begin{eqnarray}}
\newcommand{\eq}{\end{eqnarray}}
\begin{document}
\title{Classical solutions for the Lorentz-violating and CPT-even term of the
Standard Model Extension}
\author{Rodolfo Casana$^{a}$, Manoel M. Ferreira Jr$^{a}$, Carlos E. H. Santos$^{b} $}
\affiliation{$^{a}$Departamento de F\'{\i}sica, Universidade Federal do Maranh\~{a}o
(UFMA), Campus Universit\'{a}rio do Bacanga, S\~{a}o Lu\'{\i}s - MA,
65085-580, Brasil}
\affiliation{$^{b}$Departamento de F\'{\i}sica, Universidade Federal da Para\'{\i}ba
(UFPB), Caixa Postal 5008, Jo\~{a}o Pessoa - PB, 58051-970, Brasil}

\begin{abstract}
In this work, we calculate the classical solutions for the electrodynamics
stemming from the Lorentz-violating (LV)\ and CPT-even term of the standard
model extension. Static and stationary solutions for pointlike and extended
charges are obtained from the wave equations by means of the Green method. A
dipolar expansion is written for the \ field strengths. It is explicitly shown
that charge and current generate LV\ first order effects for the magnetic and
electric fields, respectively. Using the magnetic field generated by a
macroscopic $1C$ charged sphere, we establish a stringent bound for the LV
parameter: $\kappa\leq10^{-16}.$

\end{abstract}

\maketitle

\section{Introduction}

Einstein%
\'{}%
s principle of relativity sets up the Lorentz covariance as a fundamental
symmetry of physics. The establishment of this principle as a truth of nature
has been confirmed at a high level of precision by very sensitive experiments
involving resonant cavities, masers \cite{Cavity}, microwave resonators
\cite{MO}, new versions of the Michelson-Morley experiments \cite{MM}, and CPT
probing configurations \cite{CPT}. The approximate or real exactness of
Lorentz symmetry is an important issue with interesting consequences on the
Planck scale physics. Indeed, since the demonstration about the possibility of
Lorentz and CPT spontaneous breaking in the context of string theory
\cite{Samuel}, Lorentz-violating (LV) effects in the context of low-energy
physical systems have been searched as a remanent outcome of Lorentz breakdown
in the Planck scale. Such a question is of obvious interest for the
development of a quantum theory of gravity. Actually, the main theoretical
framework that governs such investigations is the standard model extension
(SME) \cite{Colladay}, which embodies Lorentz-violating coefficients in all
sectors of interaction of the usual standard model. In the context of the SME,
many authors have performed valuable contributions in several different
respects \cite{General}-\cite{Bailey}

In this work, we focus on the gauge sector of the Standard Model Extension,
whose full Lagrangian is composed of the terms,
\begin{equation}
\mathcal{L}=-\frac{1}{4}F_{\alpha\nu}F^{\alpha\nu}-\frac{1}{4}\varepsilon
_{\beta\alpha\rho\varphi}V^{\beta}A^{\alpha}F^{\rho\varphi}-\frac{1}%
{4}W_{\alpha\nu\rho\varphi}F^{\alpha\nu}F^{\rho\varphi}-J_{\alpha}A^{\alpha}.
\end{equation}
Here, the second term is the well-known CPT-odd Carroll-Field-Jackiw term,
$\varepsilon_{\beta\alpha\rho\varphi}V^{\beta}A^{\alpha}F^{\rho\varphi},$
first proposed in 1990 \cite{Jackiw}. The parameter $V^{\beta}$ stands for the
fixed background responsible for Lorentz and CPT violation and has mass
dimension $+1$. It was very strongly constrained $\left(  V^{\beta}%
\leq10^{-33}eV\right)  $ by\ birefringence data from the light of distant
astronomical systems \cite{Jackiw}. Since then the Carroll-Field-Jackiw
electrodynamics has been examined in several distinct aspects, addressing
consistency and quantization aspects \cite{Adam}, classical solutions
\cite{Classical,Casana}, Cerenkov radiation \cite{Cerenkov}, and induced
corrections to the cosmic background radiation \cite{CBR}.

On the other hand, the CPT-even term $W_{\alpha\nu\rho\varphi}F^{\alpha\nu
}F^{\rho\varphi}$ has not received the same attention, although already
examined to some extent \cite{KM1}-\cite{Bailey}. The tensor coefficient
$W_{\alpha\nu\rho\varphi}$ is dimensionless and has the same symmetries of the
Riemann tensor $\left[  W_{\alpha\nu\rho\varphi}=-W_{\nu\alpha\rho\varphi
},W_{\alpha\nu\rho\varphi}=-W_{\alpha\nu\varphi\rho},W_{\alpha\nu\rho\varphi
}=W_{\rho\varphi\alpha\nu}\right]  $ and a double null trace which yields only
19 independent components.

In the present work, we follow the prescription stated in Ref. \cite{KM1},
where the background tensor $W_{\alpha\nu\rho\varphi}$ is written in terms of
four $3\times3$ matrices $\kappa_{DE},\kappa_{DB},\kappa_{HE},\kappa_{HB},$
defined as:
\begin{equation}
\left(  \kappa_{DE}\right)  ^{jk}=-2W^{0j0k},\left(  \kappa_{HB}\right)
^{jk}=\frac{1}{2}\epsilon^{jpq}\epsilon^{klm}W^{pqlm},\left(  \kappa
_{DB}\right)  ^{jk}=-\left(  \kappa_{HE}\right)  ^{kj}=\epsilon^{kpq}W^{0jpq}.
\end{equation}
The matrices $\kappa_{DE},\kappa_{HB}$ contain together 11 independent
components while $\kappa_{DB},\kappa_{HE}$ possess together 8 components,
which sums the 19 independent elements of the tensor $W_{\alpha\nu\rho\varphi
}$. Such coefficients can be parameterized in terms of\ four tilde matrices
and one trace element, written as suitable combinations of $\kappa_{DE}%
,\kappa_{DB},\kappa_{HE},\kappa_{HB}$, namely
\begin{align}
\left(  \widetilde{\kappa}_{e+}\right)  ^{jk} &  =\frac{1}{2}(\kappa
_{DE}+\kappa_{HB})^{jk},\text{ }\left(  \widetilde{\kappa}_{e-}\right)
^{jk}=\frac{1}{2}(\kappa_{DE}-\kappa_{HB})^{jk}-\frac{1}{3}\delta^{jk}%
(\kappa_{DE})^{ii},\\
\left(  \widetilde{\kappa}_{o+}\right)  ^{jk} &  =\frac{1}{2}(\kappa
_{DB}+\kappa_{HE})^{jk},\text{ }\left(  \widetilde{\kappa}_{o-}\right)
^{jk}=-\frac{1}{2}(\kappa_{DB}-\kappa_{HE})^{jk},\left(  \widetilde{\kappa
}_{tr}\right)  ^{jk}=\frac{1}{3}(\kappa_{DE})^{ii}.
\end{align}

Ten of the 19 elements of the tensor $W_{\alpha\nu\rho\varphi}$ (belonging to
the matrices $\widetilde{\kappa}_{e+}$ and $\widetilde{\kappa}_{o-}$) are
strongly constrained (1 part in $10^{32})$ by birefringence data (see Refs.
\cite{KM1,KM2,Kob}). From the 11 independent components of the matrices
$\widetilde{\kappa}_{e+},\widetilde{\kappa}_{e-}$, five are constrained by
birefringence. For calculation purposes, we will suppose that the other six
nonbirefringent coefficients are null, which is equivalent to choosing
$\kappa_{DE}=\kappa_{HB}=0$ (including $\widetilde{\kappa}_{tr}=0$). On the
other hand, the matrices $\widetilde{\kappa}_{o-}$ and $\widetilde{\kappa
}_{o+}$ comprise eight elements, from which five are bounded by birefringence.
The three remaining coefficients are our object of investigation in this work.
The birefringence limitation over $\widetilde{\kappa}_{o-}$ can be read as%

\begin{equation}
\text{ }(\kappa_{DB}-\kappa_{HE})\leq10^{-32},
\end{equation}
which is compatible with $\kappa_{DB}=\kappa_{HE}\neq0,$ as proposed in
Ref.\cite{Kob}. The conditions $\left(  \kappa_{DB}\right)  =-\left(
\kappa_{HE}\right)  ^{T}$ and $\kappa_{DB}=\kappa_{HE}$ imply together that
the matrix $\kappa_{DB}=\kappa_{HE}$ is antisymmetric, presenting only three
non-null elements (the nonbirefringent ones). These are the only non vanishing
LV coefficients of the tensor $W_{\alpha\nu\rho\varphi}$ to be regarded from
now on, and can be expressed in terms of a vector $\kappa^{j}=\frac{1}%
{2}\epsilon^{jpq}\left(  \kappa_{DB}\right)  ^{pq}$. The present approach is
equivalent to considering $\left(  \widetilde{\kappa}_{e+}\right)
^{jk}=\left(  \widetilde{\kappa}_{e-}\right)  ^{jk}=\left(  \widetilde{\kappa
}_{o-}\right)  ^{jk}=\widetilde{\kappa}_{tr}=0,$ $\left(  \widetilde{\kappa
}_{o+}\right)  ^{jk}=\left(  \kappa_{DB}\right)  ^{jk},$ which means that we
are regarding as null the parity-even sector of $W_{\alpha\nu\rho\varphi}$
(due to the assumption $\kappa_{DE}=\kappa_{HB}=0),$ while the parity--odd
sector is reduced to three elements. The possibility of adopting different
choices of parameters, as it is discussed in Ref. \cite{Klink2}, should
mentioned. Nowadays, the $\kappa^{j}$ nonbirefringent coefficients are
constrained by microwave cavity experiments \cite{KM1}, which impose
$\kappa^{j}\leq10^{-12},$ and by the absence of vacuum Cerenkov radiation for
ultrahigh-energy cosmic rays (UHECRs) \cite{Klink1}, which state $\kappa
^{j}<10^{-17}-10^{-18}$.

In this work, we aim at evaluating the classical solutions of the Maxwell
electrodynamics supplemented by the LV $\mathbf{\boldsymbol{\kappa}-}$vector,
in an extension to the developments of Ref. \cite{Bailey}. We take as a
starting point the modified Maxwell equations and the wave equations for the
potentials and field strengths. Such equations show that charges contribute to
the magnetic sector and currents contribute to the electric field. Such
contributions are explicitly carried out by means of the Green method, which
provides solutions for pointlike and spatially extended sources. The key-point
is the expression for the scalar potential, written for a general source
$\left(  \rho,\mathbf{j}\right)  .$ From it, we obtain the electric and
magnetic field strength at second order in $\kappa.$ A dipolar expansion is
evaluated for these fields, revealing that the current contributions for the
electric field and the charge contributions for the magnetic field are first
order ones. We finalize establishing an upper bound for the LV\ parameter as
stringent as $k<10^{-16},$ a nice value for an Earth based laboratory experiment.

\section{Classical Electrodynamics stemming from the CPT-even term}

Focusing specifically on the CPT-even term $\left(  V^{\beta}=0\right)  ,$ the
Euler-Lagrange equation leads to the following motion equation:
\begin{equation}
\partial_{\nu}F^{\nu\alpha}-W^{\alpha\nu\rho\lambda}\partial_{\nu}%
F_{\rho\lambda}=J^{\alpha}, \label{c}%
\end{equation}
which contains the two modified inhomogeneous Maxwell equations, while the two
homogeneous ones come from the Bianchi identity $\left(  \partial_{\nu}%
F^{\nu\alpha\ast}=0\right)  ,$ with $F^{\alpha\beta\ast}=\frac{1}{2}%
\epsilon^{\alpha\beta\lambda\mu}F_{\lambda\mu}$ being the dual tensor. The
Maxwell equations
\begin{align}
\nabla\mathbf{\cdot E}+\mathbf{\boldsymbol{\kappa}\cdot}\left(  \nabla
\mathbf{\times B}\right)   &  =+\rho,\\
\nabla\times\mathbf{B}-\partial_{t}\left(  \mathbf{B\times\boldsymbol{\kappa}%
}\right)  -\partial_{t}\mathbf{E}+\nabla\times\left(  \mathbf{E\times
\boldsymbol{\kappa}}\right)   &  \mathbf{=}\mathbf{j,}\\
\nabla\mathbf{\cdot B}  &  \mathbf{=}0~,~\\
\nabla\times\mathbf{E}+\partial_{t}\mathbf{B}  &  \mathbf{=}\mathbf{0,}%
\end{align}
are the starting point for searching classical solutions. In order to solve
such equations, we should achieve wave equations for the vector and scalar
potentials and field strengths. Working at the stationary regime, we attain
the following equations:
\begin{align}
\nabla^{2}A_{0}\mathbf{-\boldsymbol{\kappa}}\cdot\left(  \nabla\mathbf{\times
B}\right)   &  =-\rho,\label{A0}\\
\nabla^{2}\mathbf{A+}\left[  \left(  \mathbf{\boldsymbol{\kappa}\cdot}%
\nabla\right)  \nabla-\mathbf{\boldsymbol{\kappa}}\nabla^{2}\right]  A_{0}  &
\mathbf{=}\mathbf{-}\mathbf{j.} \label{AV}%
\end{align}
Using $\nabla\cdot\mathbf{A}=0\mathbf{,}$ a consequence of the stationary
condition on the Lorentz condition $\left(  \partial_{\mu}A^{\mu}=0\right)  ,$
Eq. (\ref{A0}) takes the form:%
\begin{equation}
\nabla^{2}A_{0}\mathbf{+\boldsymbol{\kappa}}\cdot\left(  \nabla^{2}%
\mathbf{A}\right)  =-\rho. \label{A02}%
\end{equation}
The curl operator, when applied on Eq. (\ref{AV}), implies
\begin{equation}
\nabla^{2}\mathbf{B+}\left(  \mathbf{\boldsymbol{\kappa}\times}\nabla\right)
\nabla^{2}A_{0}=-\nabla\mathbf{\times j}. \label{B1}%
\end{equation}
Taking the scalar product of the vector $\mathbf{\boldsymbol{\kappa}}$ with
the entire expression (\ref{AV}) and replacing it on Eq. (\ref{A02}), we
attain a wave equation for the scalar potential, namely%
\begin{equation}
\left[  \left(  1+\mathbf{\boldsymbol{\kappa}}^{2}\right)  \nabla
^{2}\mathbf{-(\boldsymbol{\kappa}}\cdot\nabla)^{2}\right]  A_{0}%
=-\rho+\mathbf{\boldsymbol{\kappa}\cdot j}. \label{WA0}%
\end{equation}
Now, applying the full differential operator $\left[  \left(
1+\mathbf{\boldsymbol{\kappa}}^{2}\right)  \nabla^{2}%
\mathbf{-(\boldsymbol{\kappa}\cdot}\nabla)^{2}\right]  $ on Eqs. (\ref{AV})
and (\ref{B1}), it leads to intricate wave equations for the vector potential
and magnetic field strength
\begin{align}
\nabla^{2}\left[  (1+\mathbf{\boldsymbol{\kappa}}^{2})\nabla^{2}%
\mathbf{-(\boldsymbol{\kappa}\cdot}\nabla)^{2}\right]  \mathbf{A}  &
=[\left(  \mathbf{\boldsymbol{\kappa}\cdot}\nabla\right)  \nabla
-\mathbf{\boldsymbol{\kappa}}\nabla^{2}\mathbf{]}\left[  \rho
-\mathbf{\boldsymbol{\kappa}\mathbf{\cdot j}}\right]  -\mathbf{[(}%
1+\mathbf{\boldsymbol{\kappa}}^{2})\nabla^{2}\mathbf{-}\left(
\mathbf{\boldsymbol{\kappa}}\cdot\nabla\right)  ^{2}]\mathbf{j},\label{AV2}\\
\nabla^{2}\left[  (1+\mathbf{\boldsymbol{\kappa}}^{2})\nabla^{2}%
\mathbf{-(\boldsymbol{\kappa}\cdot}\nabla)^{2}\right]  \mathbf{B}  &  =\left(
\mathbf{\boldsymbol{\kappa}\times}\nabla\right)  \nabla^{2}\left[
\rho-(\mathbf{\boldsymbol{\kappa}}\cdot\mathbf{j)}\right]  -\left[
(1+\mathbf{\boldsymbol{\kappa}}^{2})\nabla^{2}\mathbf{-(\boldsymbol{\kappa}%
}\cdot\nabla)^{2}\right]  \nabla\times\mathbf{j.} \label{B2}%
\end{align}

An alternative and simpler relation for the magnetic field can be derived from
Eq. (\ref{B1})\textbf{,} which implies%
\begin{equation}
\mathbf{B=\boldsymbol{\kappa}\times E}+\frac{1}{4\pi}\nabla\times\int
d^{3}\mathbf{r}^{\prime}~\frac{\mathbf{j}\left(  \mathbf{r}^{\prime}\right)
}{\left\vert \mathbf{r-r}^{\prime}\right\vert }~. \label{B3}%
\end{equation}
which relates the magnetic field with the electric field and the current. The
term\ $-\left\vert \mathbf{r-r}^{\prime}\right\vert ^{-1}/4\pi$ is the usual
Green function of the Laplacian operator $\nabla^{2}$. It gives an easy way to
evaluate the magnetic field generated by a generic sources $\left(
\rho,\mathbf{\mathbf{j}}\right)  $, once the electric field is known.

All these wave equations reveal that the electric and magnetic sectors are
closely entwined in the sense that both charge and current generate both
magnetic and electric field strengths. Such connection in this model was
discussed in Refs. \cite{Kob,Bailey} and also appears in the case of the
Carroll-Field-Jackiw (CFJ) electrodynamics \cite{Jackiw,Casana} for a pure
spacelike background. The difference is that in the present case this
connection is really manifest in the solutions for any background
configuration, whereas the electric and magnetic CFJ solutions remain
uncoupled for the case of a purely timelike background.

A general solution for Eq. (\ref{WA0}) can be given by the integral
expression,
\begin{equation}
A_{0}\left(  \mathbf{r}\right)  =\int G(\mathbf{r-r}^{\prime})[-\rho
(\mathbf{r}^{\prime})+\mathbf{\boldsymbol{\kappa}\cdot j(r^{\prime})}%
]d^{3}\mathbf{r}^{\prime}, \label{A0S}%
\end{equation}
\ where $G(\mathbf{r-r}^{\prime})$\ is the associated Green function which
fulfills the differential equation,
\begin{equation}
\left[  (1+\mathbf{\boldsymbol{\kappa}}^{2})\nabla^{2}%
\mathbf{-(\boldsymbol{\kappa}\cdot}\nabla)^{2}\right]  G(\mathbf{r}%
-\mathbf{r}^{\prime})=\delta^{3}(\mathbf{r}-\mathbf{r}^{\prime}).
\end{equation}
\ In order to achieve $G(\mathbf{r}),$\ we use the Fourier transform:
$G(\mathbf{r}-\mathbf{r}^{\prime})=\left(  2\pi\right)  ^{-3}\int
d^{3}\mathbf{p~}\tilde{G}\left(  \mathbf{p}\right)  \exp\left[  -i(\mathbf{r}%
-\mathbf{r}^{\prime})\right]  $, so that $\tilde{G}\left(  \mathbf{p}\right)
=-[\mathbf{p}^{2}(1+\mathbf{\boldsymbol{\kappa}}^{2}\sin^{2}\alpha)]^{-1}%
,$\ with $\alpha$\ being the angle defined by the background vector $\left(
\mathbf{\boldsymbol{\kappa}}\right)  $\ and the vector $\mathbf{p},$\ so that
$\mathbf{\boldsymbol{\kappa}\cdot p}=\kappa p\cos\alpha$. \ Here, we need to
define the spherical coordinates of the momentum vector, $\mathbf{p}%
=(p,\theta,\phi),$\ and the coordinates of the fixed background,
$\mathbf{\boldsymbol{\kappa}}=(\kappa,\theta_{1},\phi_{1})$. For calculation
purposes, we align the vector\textbf{\ }$(\mathbf{r}-\mathbf{r}^{\prime})$
with the $z$-axis, so that $\theta_{1}$ is the angle defined by the vectors
$\mathbf{\boldsymbol{\kappa}}$ and $(\mathbf{r}-\mathbf{r}^{\prime})$ $\left[
\mathbf{\boldsymbol{\kappa}\cdot}(\mathbf{r}-\mathbf{r}^{\prime}%
)=\kappa\left\vert \mathbf{r-r}^{\prime}\right\vert \cos\theta_{1}\right]  $,
$\theta$ is the angle defined by the vectors $\mathbf{p}$ and $(\mathbf{r}%
-\mathbf{r}^{\prime})$ $\left[  \mathbf{p}\cdot(\mathbf{r}-\mathbf{r}^{\prime
})=p\left\vert \mathbf{r-r}^{\prime}\right\vert \cos\theta\right]  .$ In this
case, the angle $\alpha$ is given by $\cos\alpha=\cos\theta\cos\theta_{1}%
+\sin\theta\sin\theta_{1}\cos(\phi-\phi_{1}).$ The Fourier transform of
$\ \tilde{G}\left(  \mathbf{p}\right)  $ can not be solved exactly, but an
explicit solution can be achieved for the case $\mathbf{\boldsymbol{\kappa}%
}^{2}<<1,$ for which it holds $(1+\mathbf{\boldsymbol{\kappa}}^{2}\sin
^{2}\alpha)^{-1}\simeq(1-\mathbf{\boldsymbol{\kappa}}^{2}\sin^{2}\alpha).$ The
Green function then takes the form
\begin{equation}
G(\mathbf{r}-\mathbf{r}^{\prime})=-\frac{1}{4\pi}\left\{  \left(
1-\frac{\kappa^{2}}{2}\right)  \frac{1}{\left\vert \mathbf{r-r}^{\prime
}\right\vert }+\frac{\left(  \mathbf{\boldsymbol{\kappa}\cdot(r-r}^{\prime
})\right)  ^{2}}{2\left\vert \mathbf{r-r}^{\prime}\right\vert ^{3}}\right\}
\label{Green1}%
\end{equation}

Using the Green function (\ref{Green1}) and Eq. (\ref{A0S}), the scalar
potential due to general sources (at order $\mathbf{\boldsymbol{\kappa}}^{2}$)
is
\begin{equation}
A_{0}\left(  \mathbf{r}\right)  =\frac{1}{4\pi}\left\{  c(\kappa)\int
d^{3}\mathbf{r}^{\prime}~\frac{\rho\left(  \mathbf{r}^{\prime}\right)
}{\left\vert \mathbf{r-r}^{\prime}\right\vert }-\frac{1}{2}\int d^{3}%
\mathbf{r}^{\prime}~\frac{\left[  \mathbf{\boldsymbol{\kappa}\cdot}\left(
\mathbf{r-r}^{\prime}\right)  \right]  ^{2}}{\left\vert \mathbf{r-r}^{\prime
}\right\vert ^{3}}\rho\left(  \mathbf{r}^{\prime}\right)  -\int d^{3}%
\mathbf{r}^{\prime}~\frac{\mathbf{\boldsymbol{\kappa}\cdot j}\left(
\mathbf{r}^{\prime}\right)  }{\left\vert \mathbf{r-r}^{\prime}\right\vert
}\right\}  .\ \label{A03}%
\end{equation}
with $c(\kappa)=\left(  1-\kappa^{2}/2\right)  .$ Such expression reveals that
the Lorentz-violating charge corrections to $A_{0}$ are proportional to
$\mathbf{\boldsymbol{\kappa}}^{2},$ while the current corrections are of first
order in $\mathbf{\boldsymbol{\kappa}.}$ With this expression, we may
immediately evaluate the scalar potential for a pointlike charge at rest
$\left[  \rho(\mathbf{r}^{\prime})=e\delta(\mathbf{r}^{\prime})\right]  $ and
a pointlike charge at stationary motion with velocity $\mathbf{u}$, $\left[
\mathbf{j}(\mathbf{r}^{\prime})=e\mathbf{u}\delta(\mathbf{r}^{\prime})\right]
.$ Direct integration yields%

\begin{equation}
A_{0}\left(  \mathbf{r}\right)  =\frac{e}{4\pi}\left\{  \frac{c\left(
\kappa\right)  }{r}-\frac{\mathbf{\boldsymbol{\kappa}\cdot u}}{r}-\frac{1}%
{2}\frac{\left(  \mathbf{\boldsymbol{\kappa}\cdot r}\right)  ^{2}}{r^{3}%
}\right\}  ,
\end{equation}
This potential leads to the following expressions for the electric field of a
static and stationary charge:%
\begin{align}
\mathbf{E}\left(  \mathbf{r}\right)   &  =\frac{e}{4\pi}\left\{  c\left(
\kappa\right)  \frac{\mathbf{r}}{r^{3}}\mathbf{-}\frac{3\left(
\mathbf{\boldsymbol{\kappa}\cdot r}\right)  ^{2}}{2r^{5}}\mathbf{r}%
+\frac{\left(  \mathbf{\boldsymbol{\kappa}\cdot r}\right)  }{r^{3}%
}\mathbf{\boldsymbol{\kappa}}\right\}  ,\label{ES1}\\
\mathbf{E}\left(  \mathbf{r}\right)   &  =-\frac{e}{4\pi}\left(
\mathbf{\boldsymbol{\kappa}\cdot u}\right)  \frac{\mathbf{r}}{r^{3}},
\end{align}
respectively\textbf{.} Here, both fields present a $1/r^{2}$ decaying
behavior. Although the static field (\ref{ES1}) decays as $1/r^{2}$, \ its
behavior is non-Coulombian once the magnitude of the second term changes with
direction and the third term points in the $\mathbf{\boldsymbol{\kappa}-}%
$direction. The presence of the coefficient $c\left(  \kappa\right)  $ in the
Coulombian term reveals that the LV background also induces a screening in the
magnitude of the electric charge. Such effects may be contrasted with the ones
induced by the Carroll-Field-Jackiw background $\left(  V^{\beta}\right)  $ on
the Maxwell theory. Indeed, it is known that the electric field engendered by
a static (or moving stationary charge) remains exactly Coulombian for the case
of a timelike background, $V^{\beta}=($v$_{0},0)$ \cite{Casana}.

The electric field generated by the sources $\left(  \rho,\mathbf{\mathbf{j}%
}\right)  $, read off from Eq.(\ref{A03})
\begin{equation}
\mathbf{E}\left(  \mathbf{r}\right)  =\frac{1}{4\pi}\left\{  c\left(
\kappa\right)  \int\frac{\rho\left(  \mathbf{r}^{\prime}\right)  \left(
\mathbf{r-r}^{\prime}\right)  }{\left\vert \mathbf{r-r}^{\prime}\right\vert
^{3}}d^{3}\mathbf{r}^{\prime}~+\frac{1}{2}\nabla\int d^{3}\mathbf{r}^{\prime
}~\frac{\left[  \mathbf{\boldsymbol{\kappa}\cdot}\left(  \mathbf{r-r}^{\prime
}\right)  \right]  ^{2}}{\left\vert \mathbf{r-r}^{\prime}\right\vert ^{3}}%
\rho\left(  \mathbf{r}^{\prime}\right)  -\int~\frac{\mathbf{\boldsymbol{\kappa
}\cdot j}\left(  \mathbf{r}^{\prime}\right)  }{\left\vert \mathbf{r-r}%
^{\prime}\right\vert ^{3}}\left(  \mathbf{r-r}^{\prime}\right)  d^{3}%
\mathbf{r}^{\prime}\right\}  . \label{B5a0}%
\end{equation}
The magnetic field stemming from Eq.(\ref{B3}) is then\textbf{\ }given by (at
$\mathbf{\boldsymbol{\kappa}}^{2}$ order)
\begin{equation}
\mathbf{B}\left(  \mathbf{r}\right)  =\frac{1}{\left(  4\pi\right)  }\left\{
\int\frac{\rho\left(  \mathbf{r}^{\prime}\right)  \mathbf{\boldsymbol{\kappa
}\times}\left(  \mathbf{r-r}^{\prime}\right)  }{\left\vert \mathbf{r-r}%
^{\prime}\right\vert ^{3}}d^{3}\mathbf{r}^{\prime}~-\int~\frac
{\mathbf{\boldsymbol{\kappa}\times}\left(  \mathbf{r-r}^{\prime}\right)
\left(  \mathbf{\boldsymbol{\kappa}\cdot j}\left(  \mathbf{r}^{\prime}\right)
\right)  }{\left\vert \mathbf{r-r}^{\prime}\right\vert ^{3}}d^{3}%
\mathbf{r}^{\prime}+\nabla\times\int d^{3}\mathbf{r}^{\prime}~\frac
{\mathbf{j}\left(  \mathbf{r}^{\prime}\right)  }{\left\vert \mathbf{r-r}%
^{\prime}\right\vert }\right\}  . \label{B5}%
\end{equation}
This expression shows that the charges yield a first order LV\ contribution to
the magnetic field while the currents provide only a second order
contribution. The last term of the expression above is the usual contribution
of the Maxwell theory. For the case of pointlike sources $\left[
\rho(\mathbf{r}^{\prime})=q\delta(\mathbf{r}^{\prime}),\mathbf{j}%
=q\mathbf{u}\delta(\mathbf{r}^{\prime})\right]  $, the resulting magnetic
field\ (at $\mathbf{\boldsymbol{\kappa}}^{2}$ order) is:
\begin{equation}
\mathbf{\mathbf{B}}\left(  \mathbf{r}\right)  =\frac{q}{4\pi}\left\{  \left[
1-\left(  \mathbf{\boldsymbol{\kappa}\cdot u}\right)  \right]  \frac
{\mathbf{\boldsymbol{\kappa}}\times\mathbf{r}}{r^{3}}+\frac{\mathbf{u}%
\times\mathbf{r}}{r^{3}}\right\}  . \label{B6}%
\end{equation}
Such a solution has two components: one pointing in the direction
$\mathbf{\boldsymbol{\kappa}\times r}$\textbf{,} other in the direction
$\mathbf{u\times r}$\textbf{,} revealing that the magnetic field is always
orthogonal to the position vector $\mathbf{r}$\textbf{.} For a static
pointlike charge, the associated magnetic field is
\begin{equation}
\mathbf{\mathbf{B}}\left(  \mathbf{r}\right)  =\frac{q}{4\pi}\frac
{\mathbf{\boldsymbol{\kappa}}\times\mathbf{r}}{r^{3}}. \label{B9}%
\end{equation}

For a consistency issue, it should be mentioned that the result (\ref{B6}%
)\textbf{\ }can be obtained directly from Eq. (\ref{B2}) for pointlike
sources. In fact, proposing a Fourier transform expression, $B\left(
\mathbf{r}\right)  =\left(  2\pi\right)  ^{-3}\int\tilde{B}\left(
\mathbf{p}\right)  \exp\left(  -i\mathbf{p}\cdot\mathbf{r}\right)
d^{3}\mathbf{p},$ and replacing it in the full expression (\ref{B2}), we
achieve
\begin{equation}
\tilde{B}\left(  \mathbf{p}\right)  =-q\left[  i\frac
{(1-\mathbf{\boldsymbol{\kappa}\cdot u)\boldsymbol{\kappa}\times p}%
}{\mathbf{p}^{2}(1+\mathbf{\boldsymbol{\kappa}}^{2})-\left(
\mathbf{\boldsymbol{\kappa}\cdot p}\right)  ^{2}}+i\frac{\mathbf{u\times p}%
}{\mathbf{p}^{2}}\right]  ,
\end{equation}
whose Fourier transform, at order\textbf{\ }$\kappa^{2}$, provides exactly the
outcome of Eq. (\ref{B6})\textbf{. }

\subsection{The dipole approximation}

In the case of spatially distributed sources, we can work in the dipole
approximation, $|\mathbf{r}-\mathbf{r}^{\prime}|^{-1}=r^{-1}+(\mathbf{r}%
\cdot\mathbf{r}^{\prime})/r^{3}$. With it, Eq. \textbf{(}\ref{A03}\textbf{)
}provides\textbf{\ }
\begin{equation}
A_{0}\left(  \mathbf{r}\right)  =\frac{1}{4\pi}\left\{  c\left(
\kappa\right)  \frac{q}{r}-\frac{\left(  \mathbf{\boldsymbol{\kappa}\cdot
r}\right)  ^{2}}{2r^{3}}q+\left[  \frac{c\left(  \kappa\right)  }{r^{3}}%
-\frac{3\left(  \mathbf{\boldsymbol{\kappa}\cdot r}\right)  ^{2}}{2r^{5}%
}\right]  \left(  \mathbf{r\cdot p}\right)  +\frac{\left(
\mathbf{\boldsymbol{\kappa}\cdot r}\right)  }{r^{3}}\left(
\mathbf{\boldsymbol{\kappa}}\cdot\mathbf{p}\right)  -\frac{1}{r^{3}%
}\mathbf{\mathbf{r}}\cdot\left(  \mathbf{\boldsymbol{\kappa}\mathbf{\times}%
m}\right)  \right\}  . \label{A0dip}%
\end{equation}
Here, we have used $q=\int\rho(\mathbf{r}^{\prime})d^{3}\mathbf{r}^{\prime}%
\ $as the electric charge, $\mathbf{p=}\int\mathbf{r}^{\prime}\rho
(\mathbf{r}^{\prime})d^{3}\mathbf{r}^{\prime}$ as the electric dipole moment,
and $\mathbf{m=}\frac{1}{2}\int\mathbf{r}^{\prime}\times\mathbf{j(r}^{\prime
})d^{3}\mathbf{r}^{\prime}$ as the magnetic dipole moment associated with the
current $\mathbf{j}$ (considering a localized and divergenceless current density).

The corresponding electric field is obtained from \textbf{(}\ref{A03}%
\textbf{)} via $\mathbf{E}=-\nabla A_{0}$, or introducing the dipolar
approximation directly in (\ref{B5a0})
\begin{align}
\mathbf{E}\left(  \mathbf{r}\right)   &  =\frac{1}{4\pi}\left\{  c\left(
\kappa\right)  \left[  \frac{q}{r^{3}}\mathbf{r-}\frac{\mathbf{p}}{r^{3}%
}+\frac{3\left(  \mathbf{r}\cdot\mathbf{p}\right)  }{r^{5}}\mathbf{r}\right]
-\frac{3q\left(  \mathbf{\boldsymbol{\kappa}\cdot r}\right)  ^{2}}{2r^{5}%
}\mathbf{\mathbf{r}-}\frac{15\left(  \mathbf{\boldsymbol{\kappa}\cdot
r}\right)  ^{2}\left(  \mathbf{r\cdot p}\right)  }{2r^{7}}\mathbf{r+}%
\frac{3\left(  \mathbf{\boldsymbol{\kappa}\cdot r}\right)  \left(
\mathbf{\boldsymbol{\kappa}}\cdot\mathbf{p}\right)  }{r^{5}}\mathbf{r}\right.
\label{Edip}\\
&  \left.  +\frac{3\left(  \mathbf{\boldsymbol{\kappa}\cdot r}\right)  ^{2}%
}{2r^{5}}\mathbf{p}+\frac{q\left(  \mathbf{\boldsymbol{\kappa}\cdot r}\right)
}{r^{3}}\mathbf{\boldsymbol{\kappa}}+\frac{3\left(  \mathbf{\boldsymbol{\kappa
}\cdot r}\right)  \left(  \mathbf{r\cdot p}\right)  }{r^{5}}%
\mathbf{\boldsymbol{\kappa}}-\frac{\left(  \mathbf{\boldsymbol{\kappa}}%
\cdot\mathbf{p}\right)  }{r^{3}}\mathbf{\boldsymbol{\kappa}-}\frac{3\left[
\mathbf{\mathbf{r}}\cdot\left(  \mathbf{\boldsymbol{\kappa}\mathbf{\times}%
m}\right)  \right]  }{r^{5}}\mathbf{\mathbf{r+}}\frac
{\mathbf{\boldsymbol{\kappa}}\times\mathbf{m}}{r^{3}}\right\}  .\nonumber
\end{align}
It exhibits a$\ 1/r^{2}$ behavior. The first three terms (into brackets) at
zeroth order represent the usual Coulombian behavior (the ones of the usual
Maxwell theory), whereas the following nine terms represent the non-Coulombian
electric character. The last two terms are the corrections stemming from the
magnetic moment of the current. All these terms could originate new
interesting phenomena potentially observable both at microscopic (atomic) and
macroscopic levels.

In the dipole approximation, the general expression (\ref{B3}) reads as%
\begin{equation}
\mathbf{B=\boldsymbol{\kappa}\times E}+\frac{1}{4\pi}\left[  \frac{3\left(
\mathbf{m}\cdot\mathbf{r}\right)  }{r^{5}}\mathbf{r-}\frac{\mathbf{m}}{r^{3}%
}\right]  ~. \label{gen2}%
\end{equation}
with the last two terms coming from the usual Maxwell theory.

The magnetic field in the dipole approximation can be obtained directly from
Eq.(\ref{B5}) or by using the general expression (\ref{gen2}) with
(\ref{Edip}); thus it amounts to\textbf{\ }%
\begin{equation}
\mathbf{B}\left(  \mathbf{r}\right)  =\frac{1}{4\pi}\left\{  \left[  \frac
{q}{r^{3}}+\frac{3\left(  \mathbf{r}\cdot\mathbf{p}\right)  }{r^{5}}%
-\frac{~3\mathbf{\mathbf{r}}\cdot\left(  \mathbf{\boldsymbol{\kappa
}\mathbf{\times}m}\right)  }{r^{5}}\right]  \mathbf{\boldsymbol{\kappa}}%
\times\mathbf{\mathbf{r}}-\frac{\mathbf{\boldsymbol{\kappa}}\times\mathbf{p}%
}{r^{3}}+\frac{\mathbf{\boldsymbol{\kappa}}\times\left(
\mathbf{\boldsymbol{\kappa}}\times\mathbf{m}\right)  }{r^{3}}+\left[
\frac{3\left(  \mathbf{m}\cdot\mathbf{r}\right)  }{r^{5}}\mathbf{r-}%
\frac{\mathbf{m}}{r^{3}}\right]  \right\}  . \label{Bdip}%
\end{equation}
The non-Maxwellian terms are induced by the LV background. As already noticed,
the LV first order effects are induced by the charge distribution.

\subsection{Some applications}

Now, we can make some illustrative applications. We begin evaluating the
LV\ (magnetic) contribution to the scalar potential due to a circular ring of
current $\left(  I_{0}\right)  $ of radius $R$, confined in the $x-y$ plane,
described by the following current density $j(\mathbf{r}^{\prime}%
)=I_{0}\left[  \delta(\cos\theta^{\prime})\delta(r^{\prime}-R)/R\right]
\widehat{\mathbf{e}}_{\phi^{\prime}},$ with $\widehat{\mathbf{e}}%
_{\phi^{\prime}}=-\sin\phi^{\prime}\widehat{i}+\cos\phi^{\prime}\widehat{j}$.
Since the geometry is cylindrically symmetric, we may choose the observation
point in the $x-z$ plane $(\phi=0)$ for purposes of calculation. Replacing
such current density (with $\rho=0$) in Eq.(\ref{A03}), we achieve\textbf{\ }%
\begin{equation}
A_{0}\left(  \mathbf{r}\right)  =-\frac{I_{0}R}{\left(  4\pi\right)
}\mathbf{\boldsymbol{\kappa}}\cdot\left[  -\int_{0}^{2\pi}\frac{\sin
\phi^{\prime}d\phi^{\prime}}{\sqrt{a-b\cos\phi^{\prime}}}\widehat{i}+\int
_{0}^{2\pi}\frac{\cos\phi^{\prime}d\phi^{\prime}}{\sqrt{a-b\cos\phi^{\prime}}%
}\widehat{j}\right]  ,
\end{equation}
with
\begin{equation}
a=(r^{2}+R^{2})~,~b=2rR\sin\theta. \label{ab}%
\end{equation}
While the first integral is null, the second integral yields a non null
result
\begin{equation}
A_{0}\left(  \mathbf{r}\right)  =-\frac{I_{0}R}{\left(  4\pi\right)  }\frac
{4}{\sqrt{a+b}}\left[  \frac{a}{b}K(\alpha)-\frac{a+b}{b}E(\alpha)\right]
~\mathbf{\boldsymbol{\kappa}}\cdot\widehat{j}, \label{A04}%
\end{equation}
where $K$ and $E$ represent the complete elliptic functions of first and
second kind,\ respectively, with $\alpha=\sqrt{2b/(a+b)}.$ The result of
Eq.(\ref{A04}) can be expressed as an expansion of the ratio\textbf{\ }%
$(b/2a)=[rR\sin\theta/\left(  r^{2}+R^{2}\right)  ]$%
\begin{equation}
A_{0}\left(  \mathbf{r}\right)  =-\frac{1}{4\pi}\frac{mr\sin\theta}{\left(
r^{2}+R^{2}\right)  ^{3/2}}\ \left[  1+15\frac{R^{2}r^{2}\sin^{2}\theta
}{8\left(  r^{2}+R^{2}\right)  ^{2}}+...\right]  \mathbf{\boldsymbol{\kappa}%
}\cdot\widehat{j},
\end{equation}
where $m=\pi R^{2}I_{0}$ is the magnitude of the dipolar moment associated
with the current, given as $\mathbf{m=}m\widehat{k}$. Now, it is important to
show that this result can be reconciled with the dipolar expansion of
Eq.(\ref{A0dip}). Using the identification $mr\sin\theta\widehat{j}=$
$\mathbf{m}\times\mathbf{r}$ (valid for the configuration of
evaluation)$\mathbf{,}$ we rewrite the scalar potential as:
\begin{equation}
A_{0}\left(  \mathbf{r}\right)  =\frac{1}{4\pi}\frac{1}{\left(  r^{2}%
+R^{2}\right)  ^{3/2}}\ \left[  1+15\frac{R^{2}r^{2}\sin^{2}\theta}{8\left(
r^{2}+R^{2}\right)  ^{2}}+...\right]  \mathbf{r}\cdot\left(  \mathbf{m}%
\times\mathbf{\boldsymbol{\kappa}}\right)  ,
\end{equation}
which in the limit $r\gg R$ recovers the behavior predicted in Eq.(\ref{A0dip}%
), namely:
\begin{equation}
A_{0}\left(  \mathbf{r}\right)  =\frac{1}{4\pi}\ \left[  \frac{\mathbf{r}%
\cdot\left(  \mathbf{m}\times\mathbf{\boldsymbol{\kappa}}\right)  }{r^{3}%
}+\cdots\ \right]  .
\end{equation}

In this limit the associated electric field reveals a typical dipolar behavior
as well, as can be verified by simple inspection.

Another example is the magnetic field generated by a ring of radius $R$
containing charge $\left(  Q\right)  $, located at plane $x-y$, whose charge
density is read as $\rho(r^{\prime})=(Q/2\pi R^{2})\delta(r^{\prime}%
-R)\delta(\cos\theta^{\prime}).$ The magnetic field generated by such charge
is given by the first term of Eq.(\ref{B5}),%
\begin{equation}
\mathbf{B}\left(  \mathbf{r}\right)  =\frac{1}{\left(  4\pi\right)  }\frac
{Q}{2\pi R^{2}}\left[  \int\frac{\mathbf{\boldsymbol{\kappa}\times}\left(
\mathbf{r-r}^{\prime}\right)  \delta(r^{\prime}-R)\delta(\cos\theta^{\prime}%
)}{|\mathbf{r}-\mathbf{r}^{\prime}|^{3}}d^{3}r^{\prime}\right]  , \label{B8}%
\end{equation}
which implies:
\begin{equation}
\mathbf{B}\left(  \mathbf{r}\right)  =\frac{1}{\left(  4\pi\right)  }\frac
{Q}{2\pi}\left[  \left(  \mathbf{\boldsymbol{\kappa}\times r}\right)  \int
_{0}^{2\pi}\frac{1}{[a-b\cos\phi^{\prime}]^{3/2}}d\phi^{\prime}-\left(
\mathbf{\boldsymbol{\kappa}\times}\widehat{i}\right)  R\int_{0}^{2\pi}%
\frac{\cos\phi^{\prime}}{[a-b\cos\phi^{\prime}]^{3/2}}d\phi^{\prime}\right]  ,
\end{equation}
with the parameters $a,b$ given by Eq.(\ref{ab}), and it was used in
$\mathbf{R=}R(\cos\phi^{\prime}\widehat{i}+\sin\phi^{\prime}\widehat{j})$. The
corresponding solution is
\begin{equation}
\mathbf{B}\left(  \mathbf{r}\right)  =\frac{1}{\left(  4\pi\right)  }\frac
{Q}{2\pi}\frac{\mathbf{4}}{\sqrt{a+b}(a-b)}\left\{  \left(
\mathbf{\boldsymbol{\kappa}\times r}\right)  E(\alpha)-R\left(
\mathbf{\boldsymbol{\kappa}\times}\widehat{i}\right)  \left[  \frac{\left(
a-b\right)  }{b}K(\alpha)-\frac{a}{b}E(\alpha)\right]  \right\}  .
\end{equation}

An expansion in terms of the ratio $[rR\sin\theta/\left(  r^{2}+R^{2}\right)
]$ can be performed, implying
\begin{align}
\mathbf{B}\left(  \mathbf{r}\right)   &  =\frac{Q}{2\pi^{2}}\frac{1}{\left(
r^{2}+R^{2}\right)  ^{3/2}}\left\{  \left(  \mathbf{\boldsymbol{\kappa}\times
r}\right)  \frac{\pi}{2}\left[  1+\frac{15}{4}\frac{\left(  rR\sin
\theta\right)  ^{2}}{\left(  r^{2}+R^{2}\right)  ^{2}}\right]  \right.
\\[0.2cm]
&
~\ \ \ \ \ \ \ \ \ \ \ \ \ \ \ \ \ \ \ \ \ \ \ \ \ \ \ \ \ \ \ \ \ \ \ \ \ \ \ \ \ \ \ \ \ \ \left.
-R\left(  \mathbf{\boldsymbol{\kappa}\times}\widehat{i}\right)  \frac{\pi}%
{2}\left[  -\frac{3}{2}\frac{\left(  rR\sin\theta\right)  }{\left(
r^{2}+R^{2}\right)  }-\frac{105}{16}\frac{\left(  rR\sin\theta\right)  ^{3}%
}{\left(  r^{2}+R^{2}\right)  ^{3}}\right]  \right\}  .\nonumber
\end{align}
In the limit $r\gg R,$ we have
\begin{equation}
\mathbf{B}\left(  \mathbf{r}\right)  =\frac{Q}{4\pi}\frac{\left(
\mathbf{\boldsymbol{\kappa}\times r}\right)  }{r^{3}}.
\end{equation}
This outcome coincides with the the dipolar expansion (\ref{Bdip}), once
$\mathbf{p=}\int\mathbf{r}^{\prime}\rho(\mathbf{r}^{\prime})d^{3}%
\mathbf{r}^{\prime}=0$ for the charge distribution here considered.

Finally, we shall employ the expression for the magnetic field generated by a
static charge [see Eq. (\ref{B9})] to obtain an upper bound on the
Lorentz-violating vector. First, we consider the magnetic field created by the
electric charge confined in an atomic nucleus $\left(  Z\right)  $. \ For the
sodium element, $Z=11e,$ with $e=\sqrt{1/137}.$ Evaluating the magnitude of
the magnetic field, $\left\vert \mathbf{B}\left(  \mathbf{r}\right)
\right\vert =\left(  4\pi\right)  ^{-1}Z\kappa/r^{2},$ at a typical atomic
orbital distance $\left(  r=0.75\times10^{-10}~\text{m}\right)  ,$ we obtain:
$\left\vert \mathbf{B}\left(  \mathbf{r}\right)  \right\vert =10^{5}\kappa
~$(eV)$^{2}.$ Such a field obviously couples with the electron spin, amounting
to an energy contribution $\left(  \Delta E=\mathbf{\mu}_{s}\cdot
\mathbf{B}\right)  $ that may modify the spectral lines, where $\mathbf{\mu
}_{s}=g_{s}\left(  \mu_{B}/\hslash\right)  \mathbf{S}$ is the spin magnetic
momentum and $\mu_{B}$ is the Bohr magneton. Taking $g_{s}=2,$ $S=1/2,$
$\mu_{B}=1.3\times10^{-10}~$(eV)$^{-1},$ we have $\Delta E=1.3\times
10^{-5}\kappa.$ Regarding that such correction may not be larger than
$10^{-10}~$eV$,$ the following limit is attained: $\kappa<10^{-5}.$ For
heavier atoms, this limit can be improved to $\kappa<10^{-6}$.

Another case that can provide a better bound consists of a conducting sphere
of radius equal to $R$, and endowed with a large electric charge $Q$. Once the
magnetic field from a pointlike charge goes as $r^{-2}$ - according to Eq.
(\ref{B9}), a charged sphere should engender a magnetic field proportional to
$Q\kappa/r^{2}.$ A $R=0.9~$m sphere charged with $1C$ (maintained in vacuum)
generates a magnetic field at $r=1~$m equal to $\left\vert \mathbf{B}\left(
\mathbf{r}\right)  \right\vert =2\times10^{4}\kappa~$(eV)$^{2}.$ Remembering
that superconducting quantum interference devices are able to detect magnetic
field variations as small as $10^{-10}G,$ an upper limit as stringent as
$\kappa<10^{-16}$ can be, in principle, set up.

\section{Conclusion}

In this work, we have investigated the classical solutions of the
Lorentz-violating electrodynamics associated with the CPT-even term of the
gauge sector of SME. Amongst the 19 independent components of the tensor
$W^{\alpha\nu\rho\varphi}$, we have focused on three components of the
parity-odd sector of this tensor, represented by the
$\mathbf{\boldsymbol{\kappa}-}$vector . The Maxwell equations and wave
equations for the potentials and magnetic field were written. The Green method
was applied to yield the classical solutions for static and stationary
sources. In this way, general solutions for the scalar potential, electric and
magnetic fields were constructed for pointlike charge and spatially extended
sources. A dipolar expansion was written for the field strengths. It is
explicitly shown that charges generate first order effects for the magnetic
field while currents imply first order effects for electric fields. Hence, a
suitable experiment conceived to constrain the LV parameter should involve one
of these two situations. Considering the magnetic field engendered by a
macroscopic charged sphere in vacuum, a stringent bound ($\kappa<10^{-16})$
for the LV coefficient can be stated, the best result for a laboratory based
experiment to date.

In an extension of this work, the effects of the six nonbirefringent terms
(here taken as null) belonging to the parity-even sector of tensor
$W_{\alpha\nu\rho\varphi},$ and comprised by the matrices $\widetilde{\kappa
}_{e+},\widetilde{\kappa}_{e-}$ , will be investigated. These terms are
contained in the matrix $\kappa_{HB}$ and in the coefficient $\widetilde
{\kappa}_{tr}.$ We expect that the classical solutions associated with such
terms may lead to new effects and upper bounds on these parameters. This work
is now in progress.

\begin{acknowledgments}
The authors are grateful to FAPEMA (Funda\c{c}\~{a}o de Amparo \`{a} Pesquisa
do Estado do Maranh\~{a}o), to CNPq (Conselho Nacional de Desenvolvimento
Cient\'{\i}fico e Tecnol\'{o}gico), and CAPES for financial support.
\end{acknowledgments}

\end{document}